\documentclass[letterpaper,aps,twocolumn,superscriptaddress]{revtex4}
\usepackage{graphicx}
\usepackage{epsfig}
\usepackage{amsmath,amssymb,latexsym}
\usepackage{color}

\newcommand{\bfm}[1]{\mbox{\boldmath$ #1 $}}

\def\lsim{\mathrel{\rlap{\lower4pt\hbox{\hskip1pt$\sim$}}
    \raise1pt\hbox{$<$}}}

\begin{document}

\title{Cosmic Rays in a Galactic Breeze}

\author{Andrew~M.~Taylor}
\affiliation{Dublin Institute for Advanced Studies, 31 Fitzwilliam Place, Dublin 2, Ireland\\
Phone: +353 16621333 ext.337, Email: taylora@cp.dias.ie}

\author{Gwenael~Giacinti}
\affiliation{Max-Planck-Institut f\"ur Kernphysik, 
             Saupfercheckweg 1, D-69117 Heidelberg, Germany\\
Email: giacinti@mpi-hd.mpg.de}

\begin{abstract}
Motivated by the discovery of the non-thermal Fermi bubble features both below and above the
Galactic plane, we investigate a scenario in which these bubbles are formed through 
Galacto-centric outflow. Cosmic rays (CR) both diffusing and advecting within a Galactic breeze 
outflow, interacting 
with the ambient gas present, give rise to $\gamma$-ray emission, providing an
approximately flat surface brightness profile of this emission, as observed.
Applying the same outflow profile further out within the disk, the resultant effects on the
observable CR spectral properties are determined. A hardening in the spectra due to 
the competition of advective and diffusive propagation within a particular energy range 
is noted, even in the limiting case of equal CR diffusion coefficients in the disk and halo. 
It is postulated that this hardening effect may relate to the observed 
hardening feature in the CR spectrum at a rigidity of $\approx 200$\,GV.
\end{abstract}

\newcounter{pub}

\maketitle


\section{Introduction}
\label{intro}


The presence of a Galactic wind has considerable impact on
an array of topics connected to describing the Galactic \lq\lq halo\rq\rq\/ 
environment. With little knowledge about such outer regions of
the Galaxy, information provided by non-thermal probes hold 
the first clues to revealing new information about this Galactic 
frontier.

Over the past few decades, a growing body of evidence has amounted
suggesting that our Galactic center (GC) region feeds a wind. Such indications
have been provided from a broad observational energy range, from 
radio HI~\cite{Lockman:1984}, infrared~\cite{Morris:1996} to X-ray~\cite{Cheng:1996}.
Infrared observations at larger scales~\cite{Bland-Hawthorn:2003} have further 
indicated that this wind continues out to larger scales and may be
responsible for the larger out-of-plane scale structures observed.

More recently, absorption line features in the spectra of distant AGN 
have been used to probe the gas flow structure \cite{Keeney:2006au}.
The picture provided by these results indicates the presence of 
coherent gas flow, consistent with that of an outflow directed
away from the Galactic plane.
Furthermore, recent $\gamma$-ray and radio observations 
\cite{Su:2010qj,Yang:2014pia,Fermi-LAT:2014sfa,Carretti:2013sc} 
of the region above and below the GC indicate the presence of 
extended non-thermal particle populations inside bubble structures which sit above
and below the Galactic disk. The presence of these cosmic ray (CR) populations 
are indicative of outflow activity from the GC region.
The present picture, therefore, appears to indicate that both hot gas
and non-thermal particles are conveyed out from the center of the disk 
into the halo within a centrally driven Galactic wind.

With regards the velocity of the Milky Way's outflow, there are several
indicators about this from a host of independent observations. Relatively mild 
velocities $\sim 300$\,km\,s$^{-1}$ are suggested to be present in the outflow region 
close to the disk ($\sim 1-2$\,kpc) by the weakness of the X-ray features associated with the 
bubble edges \cite{Su:2010qj,Kataoka:2013tma,Fang:2014hea,Fox:2015}.
The observation of high velocity clouds in regions consistent with the bubble's
location \cite{Keeney:2006au}, motivate outflow velocities of $\sim 150$\,km\,s$^{-1}$
at distances of $\sim 4$\,kpc and $\sim 9$\,kpc away of the Galactic plane.
Further out towards the edges of the bubbles, other indications
support velocities $<100$\,km\,s$^{-1}$ in the outflow. Within such
a profile scheme, the distortion of the outflow structures seen to 
high latitudes in radio observations \cite{Carretti:2013sc} may be 
related to the motion of the Milky Way towards Andromeda, whose 
relative velocity is $\sim 50$\,km\,s$^{-1}$.

In the following, we consider the secondary signatures that CR
embedded in outflows can produce.
In Section~\ref{GC_outflow} we adopt simple descriptions for the velocity
flow in the outflow and consider the subsequent diffusive-advective motion
of CR within it. The generation of secondary signals by these CRs are
considered in an effort for simple comparisons with recent observations. 
In Section~\ref{CR_at_Earth}, the implications of the presence of Galactic
driven outflows on the CR detected at Earth are considered.
We draw our conclusions from these results in Section~\ref{conclusion}.


\section{CR and $\gamma$-rays associated with a Galactocentric Outflow}
\label{GC_outflow}


We describe the propagation of CR within an outflow from the GC 
region using the diffusion-advection equation. Denoting $\psi_{\rm CR}({\bf r},p,t)$ 
the CR density per unit of particle momentum $p$, at {\bf r}, 
\begin{eqnarray}
\frac{\partial \psi_{\rm CR}}{\partial t} &=& \bfm{\nabla}\cdot \left( \mathcal{D} \bfm{\nabla} \psi_{\rm CR} - \bfm{V}\,\psi_{\rm CR} \right)+\frac{\partial }{\partial p}\left[\frac{p}{3} (\bfm{\nabla}\cdot \bfm{V})\psi_{\rm CR} \right]\nonumber\\ 
& & -\frac{\psi_{\rm CR}}{\tau_{\rm CR}} + \mathcal{Q}_{\rm CR} \;,
\label{diff_adv}
\end{eqnarray}
where $\mathcal{Q}_{\rm CR}$ is the source term. A diffusion scattering 
length scale of $\lambda_{10\,{\rm GV}}=3\mathcal{D}_{10\,{\rm GV}}/c=0.3$\,pc is adopted. 
For CR protons, $\tau_{\rm CR}=\tau_{pp}$ is the energy 
loss time scale from $pp$ interactions, while for CR nuclei, $\tau_{\rm CR}$ 
is the interaction time scale.

Motivated by the observations discussed above, we adopt a divergence free outflow velocity profile, 
whose $z$-dependence (in a cylindrical coordinate system where the $z$-axis is perpendicular to the disk) takes the form
\begin{eqnarray}
\bfm{V} \cdot \hat{\textbf{z}} =v_{\rm max}e^{\frac{1}{2}(1-\frac{d}{z})} \times \frac{2}{1+z/d}\;,
\label{EqnWind}
\end{eqnarray}
with $v_{\rm max}=300$\,km\,s$^{-1}$ and $d=1$\,kpc.
For such an outflow velocity profile, a timescale of $\mathcal{O}(100 \, {\rm Myr})$
is required in order for the outflow to fill a region beyond the bubbles. 
As for the source of this outflow, both a past AGN outburst event 
(see e.g.~\cite{Guo:2011eg,Guo:2011ip,Barkov:2013gda}), and a starburst 
phase or a sustained outflow 
driven by star formation in the Galactic centre (e.g.~\cite{Crocker:2014fla}) have been 
proposed in the literature. Reference~\cite{Sarkar2016} claims that the present velocity 
data are not conclusive on the type of source responsible for this outflow. 
Energetically, the starburst driven outflow luminosity is estimated to be
$(1-3)\times 10^{40}$~erg~s$^{-1}$ \cite{Crocker:2014fla}. Although the present
level of AGN activity from the Galactic center (of Sgr~A*) is considerably
below this ($L_{Sgr~A*}\sim 10^{33}$~erg~s$^{-1}$), there is a growing body of evidence 
that its level in the recent past was significantly higher 
\cite{Ponti:2010rn,Terrier:2010bn}. It therefore presently seems plausible
for either energy source to be driving the outflow.
In the present work, we prefer to keep the discussion general, adopting instead the
specific velocity profile of Eq.~(\ref{EqnWind}) as the starting point in our 
calculations.

Interestingly, such a profile broadly encapsulates the velocity profile of 
a \lq\lq breeze\rq\rq\/ solution for the isothermal outflow problem 
\cite{Chamberlain:1965,Parker:1965}. 
For such a solution, the wind is launched sufficiently subsonically that it 
accelerates without becoming transonic, before decelerating after the Bondi radius. 
The actual launching mechanism of the wind is clearly of particular importance
with regards its subsequent velocity evolution with distance. The acceleration
profile we adopt is motivated by an isothermal outflow, requiring effective heating
of the gas throughout the launching zone. 

A range of wind launching and acceleration mechanisms have been considered in the literature: 
Winds driven by supernovae (see e.g.~\cite{Dubois:2007wd}) and cosmic rays 
(see e.g. the numerical simulations presented in~\cite{Hanasz:2013esa,Peters:2015yaa,Girichidis2016}) have both been considered a possibilities.
Ref.~\cite{Simpson2016} also studied the impact of CRs on the properties of the wind.

In the following, we sudy the impact of outflows on CRs, rather than studying 
the mechanisms of wind launching and acceleration. To this end, we adopt a specific 
velocity profile as an input. The breeze profiles we consider, 
Eq.~(\ref{EqnWind}), do not correspond to the wind profiles found in the literature 
for CR-driven winds, such as in References~\cite{Ptuskin1997,Breitschwerdt:2002vs,Recchia:2016ylf,Socrates:2006dv,Everett:2007dw,Samui:2009wk,Dorfi:2013cia}. 
Nonetheless, our outflow profile is motivated for the Fermi bubbles by observations. 
Whether such a breeze profile can also describe outflows at larger 
galactocentric radii is unsure at the present time, but we note that some works, 
such as Reference~\cite{Dubois:2007wd}, argue that some galaxies can 
fail to produce successful winds with $dV/dz > 0$ at all $z$, for example because of the ram 
pressure of infalling material.

The corresponding gas density profile of our breeze description -- Eq.~(\ref{EqnWind}), plateaus within 
the decelerating flow phase. This motivates our naive constant density description 
for gas in the halo. 
For breeze outflow scenarios, the peak velocity distance depends on how deep within
the gravitational potential the wind is launched and the isothermal temperature
of the gas. Adopting motivated numbers for the mass $M$ within the bulge around 1\,kpc~\cite{Dwek:1995} and the isothermal temperature at the base of the wind~\cite{Kataoka:2013tma}, the Bondi radius~\cite{Bondi:1952} is,
$
d = 2GM/v_{\rm th}^{2}
\approx 2 \, \left(\frac{M}{10^{10}~M_{\odot}}\right)\left(\frac{300\,{\rm eV}}{kT}\right)\,{\rm kpc}.
$

\begin{figure*}[t]
\includegraphics[angle=-90,width=0.45\linewidth]{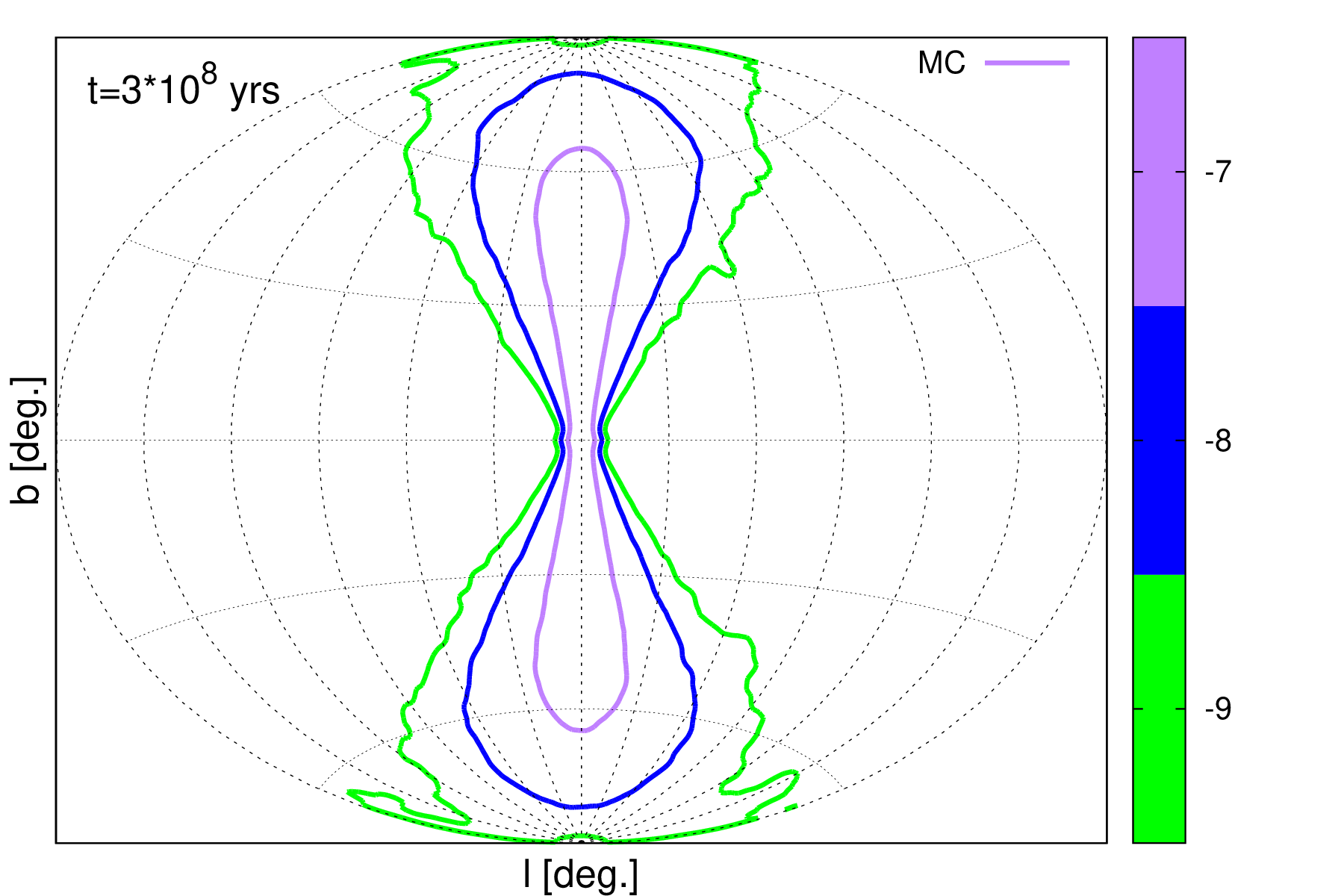}
\includegraphics[angle=-90,width=0.45\linewidth]{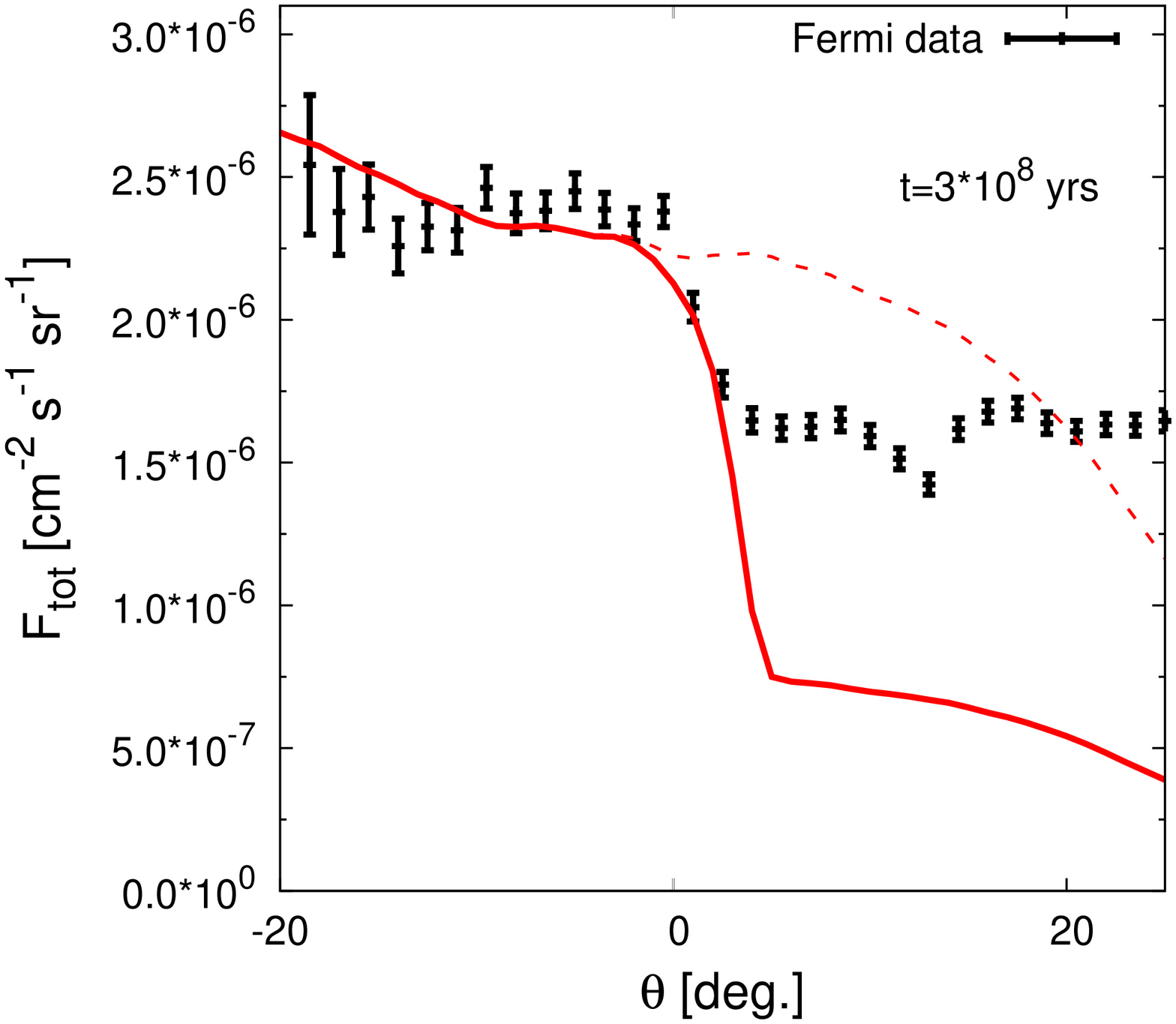}
\caption{
{\it Left:} Contour plots showing $\log_{10}$ of the $\gamma$-ray flux surface brightness (cm$^{-2}$\,s$^{-1}$\,sr$^{-1}$) from the bubbles following the interaction of CR in the outflow with the gas present. The different line colours indicate the corresponding contour value, whose values are provided in the colour bar in the side-panel. {\it Right:} A comparison of the edge of the $1-2$\,GeV $\gamma$-ray bubble from our outflow model with that from the Fermi observation analysis~\cite{Yang}. The angle $\theta$ is counted from the edge of the bubble. It is noted that for the energy bin considered, at large $\theta$, further diffuse $\gamma$-ray background \cite{Abdo:2010nz} dominates the observed flux, with the model values sitting below this level in this region. The solid line result adopts a decrease in the gas density at the bubble edge whereas the dashed line result assumes a constant density throughout.}
\label{gamma_contours}
\end{figure*}

%
%

We utilise a Monte Carlo approach to solve (\ref{diff_adv}). Our results 
with this technique have also been compared with those obtained using
an explicit differential equation solver, finding excellent agreement
in all cases (see Fig.~\ref{outflow_test} in Appendix~A).

We assume that our source term, $\mathcal{Q}_{CR}$, is constant in time and
located at the GC region. The copresence of the resultant accumulated CR 
with ambient gas gives rise to $\gamma$-ray bubble 
emission through $\pi^{0}$ production generated in $pp$ interactions.
This emission may potentially account for the observed $\gamma$-rays from 
the bubbles, as has previously been proposed by others~\cite{Crocker:2010}.

To determine the level of this emission, the accumulated CR density 
throughout the outflow region are convolved with the gas target material
density in the outflow region. As motivated on theoretical grounds by 
\cite{Feldmann:2012rx}, and on observational grounds by~\cite{Fang:2014hea}, 
we adopt a constant gas density within the bubble region at the level 
$3\times 10^{-3}$\,cm$^{-3}$. A $\gamma$-ray density map and a comparison of the 
$\gamma$-ray bubble-edge profile with Fermi measurement~\cite{Yang} 
are shown in Fig.~\ref{gamma_contours}. For these results, a CR luminosity 
of $10^{40}$\,erg\,s$^{-1}$ has been adopted for the central source. In this
comparison plot, the origin of the diffuse $\gamma$-ray emission in the $\theta>0$
region is assumed purely galactic in origin. Should some component of the
emission from this region be extragalactic, however, a subsequent reduction
of the Galactic center luminosity or bubble gas density would be required in
order to account for such a reduction in required $\gamma$-ray emission intensity.

As can be seen from Fig.~\ref{gamma_contours}, a flat surface brightness
profile for the bubbles is obtained following the assumption that
the velocity profile of the bubbles is described by Eq.~(\ref{EqnWind}).
We note though that in reality a range of velocity profiles 
can provide such a uniform brightness. See for example~\cite{Sarkar:2015xta}. 
In general, we find that for the case of a constant
density ambient medium description, the present $\gamma$-ray data can 
be said to prefer decelerating profiles. Instead, for decreasing gas density 
profiles, a sharper fall-off in the velocity profile, than that 
adopted in Eq.~(\ref{EqnWind}), would be required.

Although the cutoff at the bubble edges is not well described by
the simple constant density gas model (see dashed line in Fig.~\ref{gamma_contours}), 
a steeper cutoff in the $\gamma$-ray profile can be achieved by 
a sudden change in the density of the gas at the 
bubble edge (see solid line), as motivated in certain models~\cite{Crocker:2014fla}. 
A further motivation for such an origin for the bubble edges 
comes from a comparison of their morphology as seen 
in $\gamma$-rays~\cite{Su:2010qj} and in radio~\cite{Carretti:2013sc}.
If GeV protons and electrons respectively give rise to the $\gamma$-ray and 
radio emissions, it would be curious that the 
electrons extend out to larger latitudes than 
the protons. Such a difference between $\gamma$-ray and radio data morphologies 
disfavours simple leptonic scenarios for the $\gamma$-ray bubbles. 
Despite such challenges, however, more involved diffuse acceleration models 
supporting a scenario in which both the radio and $\gamma$-ray emission are leptonic 
in origin are also presently viable~\cite{Mertsch:2011es}.

One simple explanation for the difference 
in latitudinal profiles in the radio and $\gamma$-ray emission is that both 
protons and electrons possess extended distributions, and that the difference in morphology 
of the secondary emission they produce is dictated by differing distributions
of target gas and magnetic fields. A potential association of the astrophysical 
neutrino events detected by IceCube~\cite{Aartsen:2013jdh}, with the bubbles and 
beyond~\cite{Taylor:2014hya}, allows such a hadronic origin scenario for the 
$\gamma$-rays to be tested in the near future.


\section{Potential Implications for CR Fluxes at Earth}
\label{CR_at_Earth}

Out at radii well beyond the GC region, the role played
by any advective transport effects is less clear. In order to keep
this discussion general, we here explore two extremum cases, namely,
a diffusive only transport scenario, and a case in which the 
inferred GC outflow properties are mirrored at much
larger radii.

\subsection{Local Contamination of Galacto-centric Outflow}
\label{local_contamination}

With little evidence that a Galactic wind of an appreciable strength exists 
out at larger Galacto-centric radii, $r$, we here impose the extreme assumption that CR 
propagation in this region ($r>200$\,pc) is purely diffusive. Assuming further that the source is 
steady on the time-scales under consideration ($\mathcal{O}(100 \, {\rm Myr})$), the 
subsequent CR density along the disk is expected to follow a $1/r$ dependence 
within the region where the steady-state has been achieved, with a steeper fall 
off beyond this point.

Furthermore, TeV $\gamma$-ray observations of the GC region by the HESS 
Cherenkov telescope instrument~\cite{::2016dhk}, allow the radial distance at which the 
inferred CR density drops below its locally measured value to be determined. At 
an energy of 10\,TeV, the CR density at a distance of 100\,pc from Sgr~A* is
$\sim 6$ times above the sea level. With a $1/r$ CR density distribution, the transition 
distance is therefore $\sim 0.6$\,kpc. However, with a hard CR spectrum 
observed to be present within this region, this transition distance would be 
expected to occur at larger radii for higher energy CR. Assuming the CR energy 
density in the GC region has a spectrum $dN/dE_{\rm CR} \propto E^{-2.4}$, 
the ratio of the GC CR energy density to the sea level would be 
expected to increase as $U_{\rm GC}/U_{\rm sea}\propto E_{\rm CR}^{0.3}$. Consequently, 
assuming this scaling rule holds, a transition distance of 8\,kpc would be reached 
at an energy of $\sim 20$\,PeV. This number is derived in the most favourable case 
of no CR advection in the halo, and, therefore, should be considered as an upper limit for 
such a GC contamination. 
This shows that the GC can, in principle, contribute to 
the CR flux at PeV energies. 
However, the observed CR spectrum above the \lq\lq knee\rq\rq\/ at these energies is not 
$\propto E^{-2.4}$, implying the need for a break to exist in the spectrum. Such a
solution appears rather {\it ad hoc}, requiring fine-tuning in order that the Galactic
centre contributes to the arriving flux without, at the same time, violating spectral
shape constraints.

The PeV CR anisotropy direction, however, is compatible with a CR gradient pointing towards 
the GC. For a CR density $\propto r^{-1}$, the anisotropy amplitude is 
$\sim \lambda_{1\,{\rm PV}}/r_{\rm GC} \approx 0.6$\,\% if $\mathcal{D}\propto E^{1/3}$, which 
is close to IceCube/IceTop measurements~\cite{Aartsen:2016ivj}. We note that in a 
scenario where the PeV CR anisotropy would be due to the GC, the anisotropy below 100\,TeV 
must have a different origin. Indeed, the direction of the CR anisotropy flips 
by approximately $180^{\circ}$ around $\approx 100$\,TeV~\cite{Aartsen:2016ivj}, and 
points in the direction opposite to the GC at low energies. The anisotropy 
below 100\,TeV may, for example, be due to a nearby supernova remnant (SNR). 
Reference~\cite{Ahlers:2016njd} suggested that Vela SNR is a good candidate for shaping the CR 
anisotropy at Earth below 100\,TeV.


\subsection{Local Outflow Effects}
\label{local_outflow}


We next study the impact on CR observables of a {\em local} outflow, whose velocity gradient becomes negative above a given height $d$ in the halo. 
To our knowledge, the impact of such breeze velocity profiles on the local CR observables has not been presented in the literature yet. While such velocity profiles may not correspond to those expected for CR-driven winds (see e.g.~\cite{Breitschwerdt:2002vs,Recchia:2016ylf} where $dV/dz > 0$ at all $z$), they can be motivated in some models (see e.g. the simulations of Ref.~\cite{Dubois:2007wd} for galaxies failing to produce winds).

As a first approximation, we ignore here any variation of CR sources or propagation parameters in the radial direction from the GC axis. We assume that a one dimensional model is able to encapsulate CR propagation in the halo. 
Numerically solving the planar diffusion-advection equation in $z$ and $E$, for any arbitrary profiles of $V(z)$ and $\mathcal{D}(E)$, we coarsely investigate the effect of the advection velocity profile on CR observables. We verified that our code accurately reproduces the expected CR density profiles in the halo for the known cases of $V=$~cst~\cite{OwensJokipii1977} and $V \propto z$~\cite{Bloemen1993}, which are constant and decreasing with $z$, respectively. On the contrary, the $V(z)$ profiles we consider below lead to an increase of CR density above $d$, decreasing again as $z \rightarrow H$ (escape), where $H$ denotes the size of the escape boundary. Physically, the existence of $H$ may correspond to the height at which the magnetic field becomes too weak to confine CR through diffusion. We set $\psi = 0$ at $z=H$ as a boundary condition. We subsequently determine the steady-state distributions for $\psi_{\rm CR}(z,E)$, for protons, boron and carbon nuclei. For boron and carbon, we denote them as $\psi_{\rm B}(z,E)$ and $\psi_{\rm C}(z,E)$.

For the primary source term, we adopt the prescription: 
$Q_{\rm A} = 0$ in the halo ($|z|> h=200$\,pc), and $Q_{\rm A} = f_{\rm A}\,Q_{\rm CR}$ in the disk ($|z|\leq h$), where $f_{\rm A}$ is the fraction of species $A$ emitted at the source. For the gas density, we adopt: 
$n = 0.85$\,cm$^{-3}$ at $|z|\leq h$, and $n \sim 10^{-3}$\,cm$^{-3}$ at $|z|>h$. 
For clarity, we assume that there are no sources of primary boron.

$\psi_{\rm B,C}$ satisfy Equation~(\ref{diff_adv}), with the loss terms for 
C, N, and O, acting as source terms for boron. The source term ($\mathcal{Q}_{\rm CR}$) 
in the equation for $\psi_{\rm B}$ is:
\begin{eqnarray}
  \mathcal{Q}_{\rm B} = \sum_{\rm Z} \frac{\psi_{\rm Z}}{\tau_{\rm Z \rightarrow B}} = \sum_{\rm Z} c\sigma_{\rm Z \rightarrow B}n\psi_{\rm Z}\;,
\end{eqnarray}
where $\tau_{\rm Z \rightarrow B}=1/c\sigma_{\rm Z \rightarrow B}n$ and the contributions from nuclei $Z$ are dominated by C, N and O. Using the relative abundances of nuclei in the CR flux (see Fig.~6 of~\cite{Genolini:2015cta}), and the production cross-sections $\sigma_{\rm Z \rightarrow B}$ from~\cite{Webber2003} and quoted in Table~2 of~\cite{Genolini:2015cta}, we rewrite the boron production term as:
\begin{eqnarray}
  \mathcal{Q}_{\rm B} = \frac{\psi_{\rm C}}{\tau_{\rm \rightarrow B}} = c\sigma_{\rm \rightarrow B}n\psi_{\rm C} \;,
\end{eqnarray}
with $\tau_{\rm \rightarrow B}=1/c\sigma_{\rm \rightarrow B}n$ and $\sigma_{\rm \rightarrow B} \approx 131$\,mb. We take into account destruction of Boron (and similarly for other species) through spallation, with the decay term \lq\lq $-\psi_{\rm CR}/\tau_{\rm CR}$\rq\rq\/ in Eq.~(\ref{diff_adv}) for $\psi_{\rm B}$ set to:
\begin{eqnarray}
  -\frac{\psi_{\rm B}}{\tau_{\rm B \rightarrow}} = -c\sigma_{\rm B \rightarrow}n\psi_{\rm B} \;,
\end{eqnarray}
where $\tau_{\rm B \rightarrow}=1/c\sigma_{\rm B \rightarrow}n$, and the cross section $\sigma_{\rm B \rightarrow}$ for this process is taken from~\cite{Tripathi1999}. We find $\sigma_{\rm B \rightarrow} \approx 250$\,mb on pure p target ($\approx 276$\,mb on 90\% p + 10\% He). For clarity, we take 250\,mb, which is compatible with the value quoted in~\cite{Longair}.

\begin{figure}[t]
\includegraphics[angle=-90,width=0.49\textwidth]{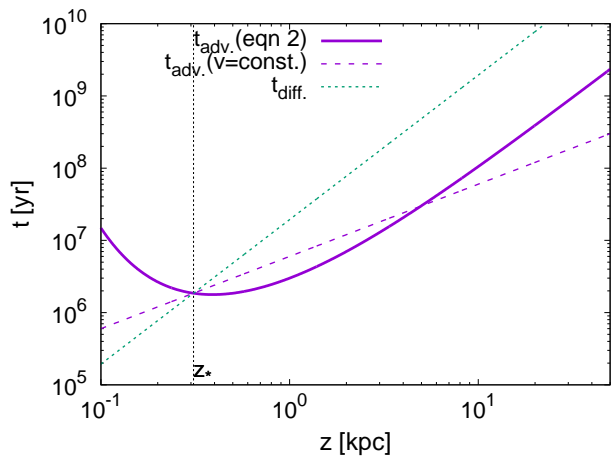}
\caption{A comparison of the transport times to different heights, $z$, above the Galactic plane for the both diffusive ($t_{\rm diff}$) and advective ($t_{\rm adv}$) transport cases. $t_{\rm diff}$ and $z_{*}$ are shown here for 10\,GV CR. The solid (dashed) line is the advection time out of region $z$ for a velocity profile described by Eq.~(\ref{EqnWind}) (for a constant velocity value), and the dotted line is the diffusion time out of the same region.}
\label{z_star}
\end{figure}

We take $H=25~$kpc, and express $\mathcal{D}$ as $\mathcal{D} = \mathcal{D}_{10\,{\rm GV}}\,(E/(Z \times 10\,{\rm GV}))^{\delta}$, setting $\delta=1/3$ and keeping the same normalization $\mathcal{D}_{10\,{\rm GV}}$ as in Section~\ref{GC_outflow}. We have verified that our code reproduces the expected B/C both for the \lq\lq no wind\rq\rq, and \lq\lq constant wind\rq\rq\/ cases. 
In the latter case, the key parameter is $z_{*}=\mathcal{D}/V$ 
(see purple dashed line in Fig.~\ref{z_star}), which separates
out the distances at which diffusion and advection dominate the
particle transport. For low energies, $z_{*}<H$ and 
particles advect to the boundary. The B/C ratio shows a 
quick transition to a constant value at these low energies because $z_{*} \propto E^{\delta}$. 
At higher energies, diffusion to the boundary begins to dominate. Since no 
sudden change of slope is seen in the B/C data, the propagation mode of CR 
in the energy range sensitive to by present experiments should be predominantly diffusive, 
i.e. $z_{*}/H>1$, demanding an advection wind speed of less than $\mathcal{O}(10 \, {\rm km \, s}^{-1})$ 
for $H\sim 10$\,kpc, in the case $V=$~cst.

In general, however, strong winds are not disallowed by the data. 
Several other wind profiles with $V\neq$~cst are not ruled out, such as $V(z)\propto z$. 
We refer to this scenario as a \lq\lq Bloemen-like\rq\rq\/ wind~\cite{Bloemen1993}. 
For such a wind, the advection time is independent of $z$, resulting 
in $z_{*}\propto \mathcal{D}^{1/2}$, and the spectral slope tending to $-\alpha-\delta$/2, when $z_{*}<H$.
Thus, the presence of such a wind would lead to a softening spectral index, 
from $-\alpha-\delta/2$ at low energies, to $-\alpha-\delta$ at high energies. Such a
profile, however, would not induce any hardening in the CR spectrum. We show now that hardenings 
can appear with more complicated wind profiles, and notably with our breeze profile, Eq.~(\ref{EqnWind}).

\begin{figure*}[t]
\includegraphics[width=0.32\textwidth]{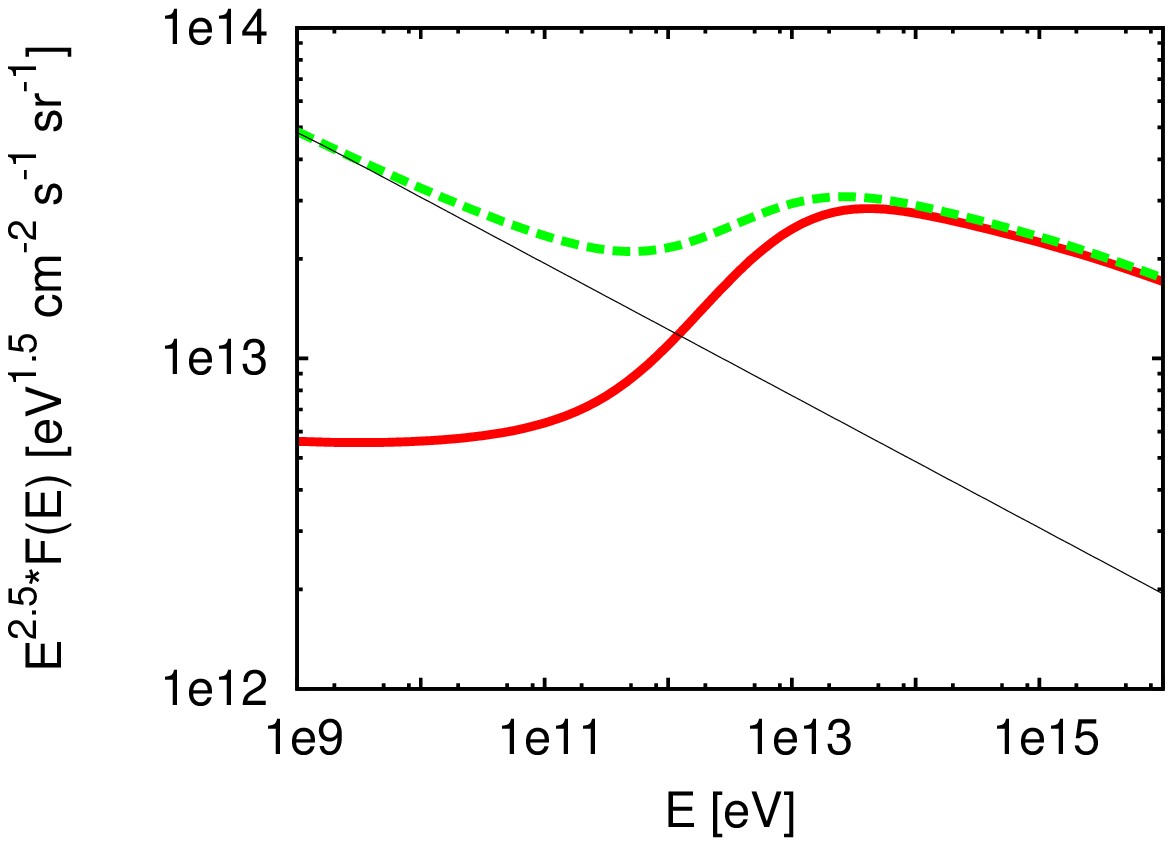}
\includegraphics[width=0.32\textwidth]{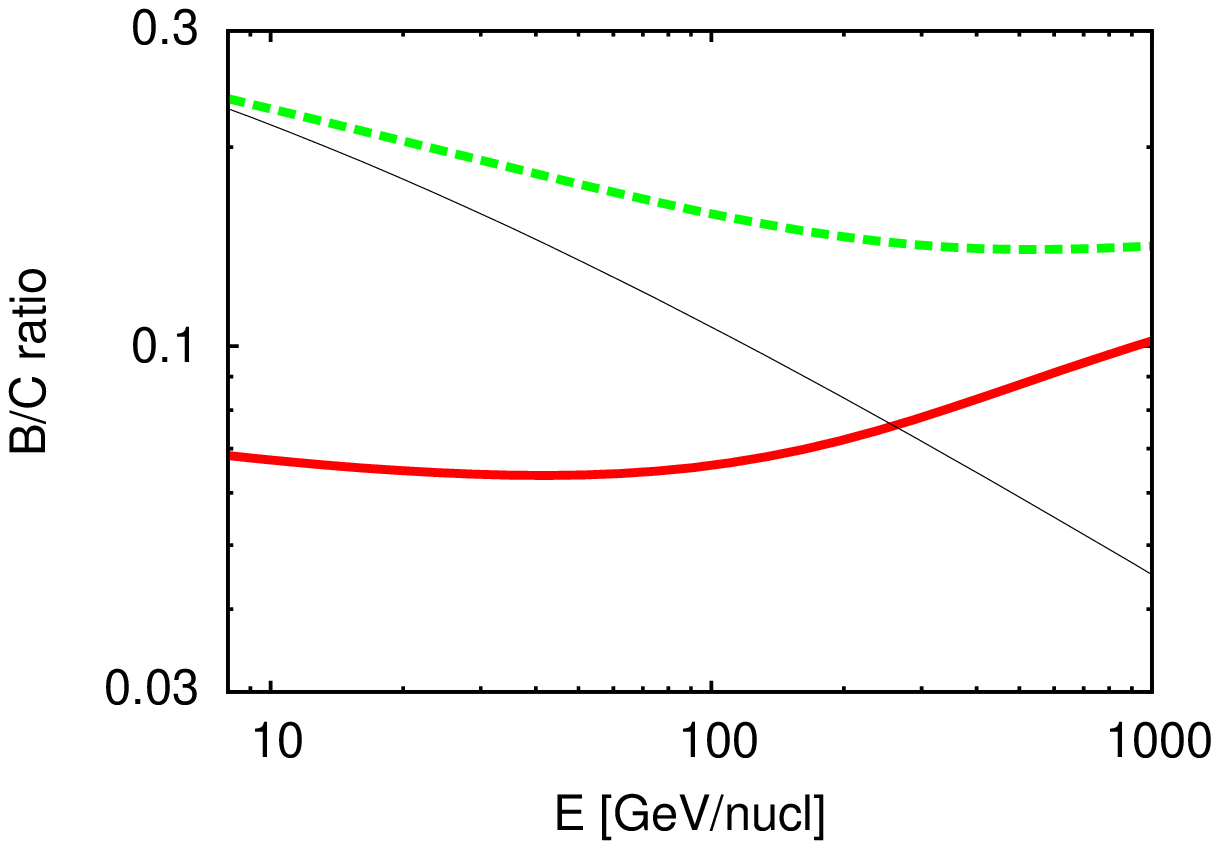}
\includegraphics[width=0.32\textwidth]{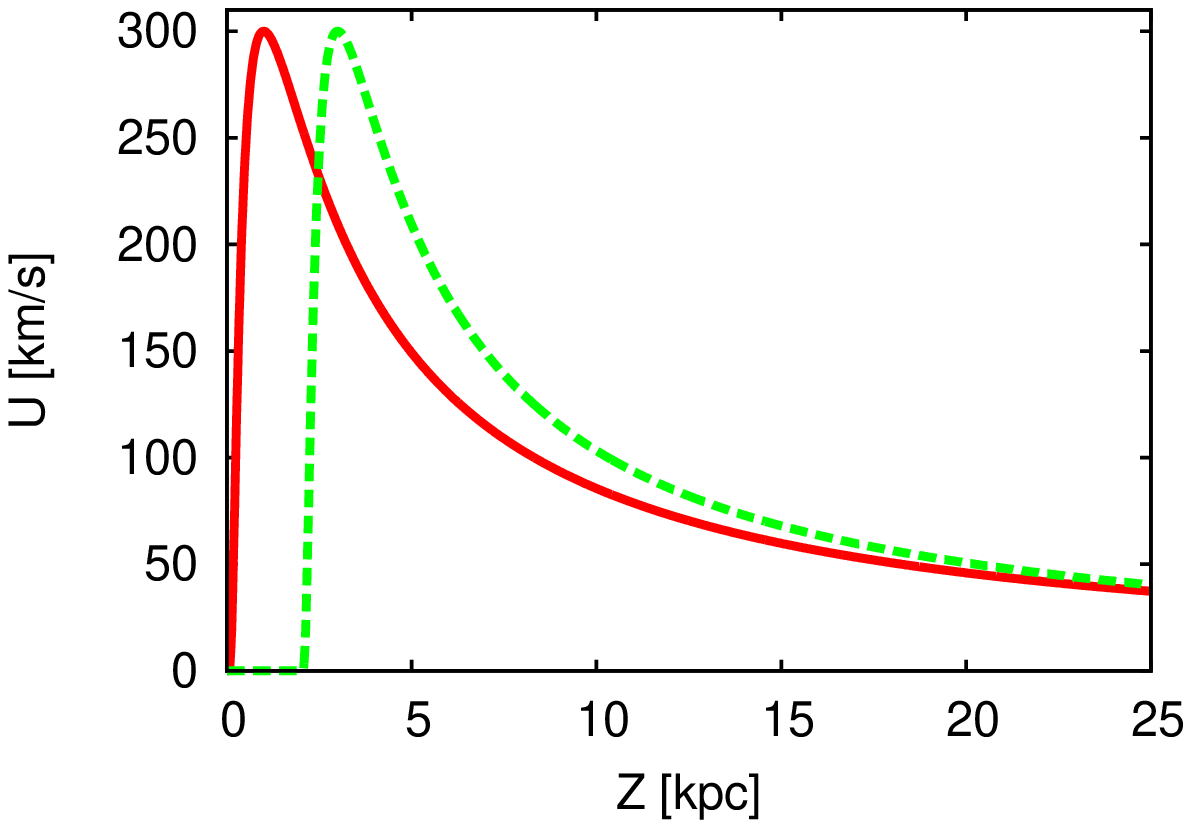}
\caption{CR flux ({\it left panel}) and B/C ratio ({\it middle}) at $z=0$, for the outflow profiles displayed in the {\it right panel}. We do {\em not} try to fit the data. $\mathcal{D}_{10\,{\rm GV}} = 3 \times 10^{28}\,{\rm cm}^2\,{\rm s}^{-1}$, $\delta = 1/3$, $H=25$\,kpc, $n(z)=0.85$\,cm$^{-3}$ for $|z| \leq h$ and $10^{-3}$\,cm$^{-3}$ otherwise, CR spectrum at sources $\propto E^{-2.37}$, and total power injected in CRs at $|z| \leq h$ set to $\approx 3.3 \times 10^{39}$\,erg\,pc$^{-2}$\,yr$^{-1}$. Each scenario is represented by the same line type on each panel. Thin black line for the \lq\lq best fit\rq\rq\/ model of~\cite{Genolini:2015cta} with $V=0$.}
\label{WindProfiles}
\end{figure*}

In Fig.~\ref{WindProfiles}, we calculate the CR spectrum at $z=0$ (left-panel), 
and the B/C ratio (middle-panel) for $V(z)$ from Eq.~(\ref{EqnWind}) (red curves), 
and for a similar profile, namely (\ref{EqnWind}) with $z \rightarrow z-2$\,kpc 
and $V=0$ at $|z|<2$\,kpc (green curves). Plots of $V(z)$ are shown in the right panel. 
For reference, we show with thin black lines the \lq\lq best fit\rq\rq\/ of the B/C ratio 
from~\cite{Genolini:2015cta} for $V=0$. The observational data from AMS-02 for the B/C ratio 
therefore coincides with this thin black line. 
We note that we do {\em not} try to fit the data. Instead, we study, on purpose, caricatural examples 
in order to explore interesting phenomena allowed by diffusion within a 
breeze profile, such as the formation of breaks or points of inflection. 
The parameter values chosen for the breeze profiles presented in Fig.~\ref{WindProfiles} make 
these features more prominent and more visible than in the data. Fitting 
the existing data will be investigated in a future work.

Focusing on the shape of the CR spectrum, one can see a point of inflection in each of
the curves. In order to interpret these inflection points, a comparison of the advection 
and diffusion timescales at different energies must be made.
In Fig.~\ref{z_star}, the advection time for particles at different heights 
above the Galactic plane are shown (solid purple line) for the case corresponding to the red solid 
line of Fig.~\ref{WindProfiles}. Continuing with the assumption that the diffusion
coefficient depends only on energy, the curvature of the $t_{\rm adv}$ curve introduces
new possibilities as to which of these transport processes dominates. This variety of
scenarios, in turn, allows for a broader range of spectral phenomena than the 
simple leaky box, $V=$~cst or \lq\lq Bloemen-like\rq\rq\/ wind descriptions. In Fig.~\ref{z_star}, 
the curve for the typical diffusion time $t_{\rm diff}$ (dotted blue line) 
crosses the advection curve, for sufficiently low energy CR (results shown here for 10\,GV CR). 
This crossing acts as a bottleneck, providing an effective halo height $z_{*}<H$. 
At low energies, the outflow then reduces the size of the diffusion \lq\lq box\rq\rq, 
within which CR can safely diffuse and return to the observer at $z=0$, from the full size, $H$, down to 
$\sim z_{*}$. Beyond this distance, advection wins over diffusion, and CR do not come back to $z=0$. 
For higher energy CR, however, the diffusion lengths are considerably larger, allowing
diffusion to win over advection in the entire halo, and the problem simplifies 
to a basic leaky-box of size $H$. With the above parameter values, this happens at energies 
$E \gtrsim 10^{13-14}$\,eV. As can be seen in Fig.~\ref{WindProfiles} (left panel), 
the CR spectrum then returns to a power-law of the form $\propto E^{-\alpha - \delta = -2.7}$ at such energies. 
Below $\sim 10^{12-13}$\,eV, the CR flux is \lq\lq suppressed\rq\rq.

A more quantitative description of this behaviour is provided through the consideration of 
the change of $z_{*}$ with diffusion coefficient, described through the relationship,
$z_{*}\propto D^{\beta}$. 
The $t_{\rm adv}$ curve in Fig.~\ref{z_star} does not vary strongly with $z$, on $0.2 \lesssim |z|/{\rm kpc} \lesssim 2$. 
Therefore, for low energy CR ($\sim 10^{9-11}$\,eV), $\beta\sim 0.5$. The resultant spectrum is then close to that of a \lq\lq Bloemen-like\rq\rq\/ wind, 
explaining why the spectral index of the red curve in Fig.~\ref{WindProfiles} (left panel) is harder than 2.7 at such energies. 
At higher energies, $\beta$ grows larger than $1$ and the role of the advection term subsequently quickly turns off. 
The effective box size abruptly increases from $z_{*} \lesssim$~a few kpc to $H$, and the resultant 
CR spectrum then becomes harder before softening again and matching the spectrum 
expected for a standard fixed-size diffusion \lq\lq box\rq\rq, at $E \gtrsim 10^{13-14}$\,eV. 
Due to this change in box size, the spectrum at high energies is normalised to a larger flux value than the spectrum at low energies. 
For the green curve, $V=0$ (i.e. $t_{\rm adv} \rightarrow \infty$) at $|z|<2$\,kpc. The corresponding $t_{\rm adv}$ 
curve would be shifted by $\approx + \,2$\,kpc at low $z$ compared to the curve shown in Fig.~\ref{z_star}. 
In this case, $t_{\rm diff}$ then crosses $t_{\rm adv}$ at a value $z_{*} \approx 2$\,kpc, 
for CR with $E \lesssim 10^{11}$\,eV. In this energy range, the increase of $z_{*}$ with 
energy is small compared to 2\,kpc. For $z$ slightly greater than 2\,kpc, 
the advection time decreases quickly with $z$, resulting in $\beta$ being small ($\ll 1$). 
The CR spectrum in this low energy region reflects that of the 
fixed-size diffusion \lq\lq box\rq\rq\/ case, with a box size equal to $\approx 2$\,kpc. This is why the spectral index of 
the green curve in Fig.~\ref{WindProfiles} (left panel) tends to 2.7 at low energies.

In summary, the spatially dependent velocity profile we adopt introduces the possibility 
for a smooth transition from one size diffusion box at low energies, to a larger 
diffusion box size at higher energies. For some parameter values, it is possible to make the 
hardening that we found in the CR spectrum coincide better with the one measured at 200\,GV 
by PAMELA, CREAM and AMS-02. Interestingly, if the high-energy softening is left concealed to higher energies ($\gtrsim 3$\,PeV), 
one may then explain the 200\,GV hardening with the launching of a breeze or wind in the halo, even {\em without} 
invoking a change in $\mathcal{D}$ between the disk and the halo. 
This argument remains valid also for some winds with $dV/dz > 0$ at all $z$. 
These data are most sensitive to the accelerating part of the outflow, while those in Sect.~\ref{GC_outflow} essentially 
probe the decelerating part of the outflow.

For the same reasons, similar hardenings are expected to appear in the B/C ratio at \lq\lq intermediate\rq\rq\/ energies, see middle panel. 
This is not contradictory with present measurements as long as the hardening is left concealed to higher energies or is hidden within the systematics of the
present instruments. Indeed, in connection to the second of these possibilities, it is noted that apparent conflicts in 
secondary to primary ratios still exist in current data sets (e.g. see Ti/Fe ratio by 
ATIC-2~\cite{Zatsepin:2009}, HEAO-3-C3~\cite{Vylet:1990}, and also their comparison to the B/C ratio~\cite{AMS2013}).


\section{Conclusions}
\label{conclusion}


We first investigated a scenario in which an advective outflow, emanating
from the Galactic center region, carries pre-accelerated CR. 
These CR produce secondary $\gamma$-rays via $pp$ interactions on target gas. 
We have demonstrated that one can reproduce a flat $\gamma$-ray surface brightness profile, 
as is observed for the Fermi bubbles, 
provided that the outflow decelerates with distance above the Galactic disk. Such a 
description for the outflow profile is 
encapsulated by \lq\lq breeze\rq\rq\/ solutions of isothermal winds.

Assuming CR propagation beyond the central zone is purely diffusive, it 
is possible for a non-negligible fraction of CR from the GC 
region to reach large radii. The contamination under this 
assumption is energy dependent, and we found that CR from
the GC may potentially become the dominant source for
the flux observed at Earth, at $\gtrsim$~PeV energies. The 
absence of evidence indicating the onset of a new component in the CR 
spectrum at these energies, however, place challenges on such a 
possibility.

Imposing, instead, a wind scenario also out at larger Galactocentric radii, we 
have demonstrated that
for the breeze profile~(\ref{EqnWind}), an inflection point
is introduced into the CR spectrum shape at $z=0$, as a result of 
competition between CR advection and diffusion in the halo. 
We have shown that hardenings can appear in the CR spectrum due to the launching 
of a breeze or wind in the Milky Way's halo, even without invoking any change in the 
CR diffusion coefficient value between the disk and halo.

We conclude that a breeze outflow scenario from the Galaxy 
provides an interesting array of observational signatures able to diagnose its
presence. Although presently only motivated from Galactocentric outflow
observations, the results outlined provide a useful reference for 
future observations able to disclose its presence at larger radii.

\section*{Acknowledgments}
AT acknowledges a Schroedinger fellowship at DIAS. The authors would
like to thank Ruizhi Yang for helpful discussions about the Fermi
data and results. Insightful discussions on the role played by the
outflow with Roland Crocker are also gratefully acknowledged.

\begin{appendix}

\section{Galactic Center Outflow Simulations}

\begin{figure}[t]
\includegraphics[angle=-90,width=0.35\textwidth]{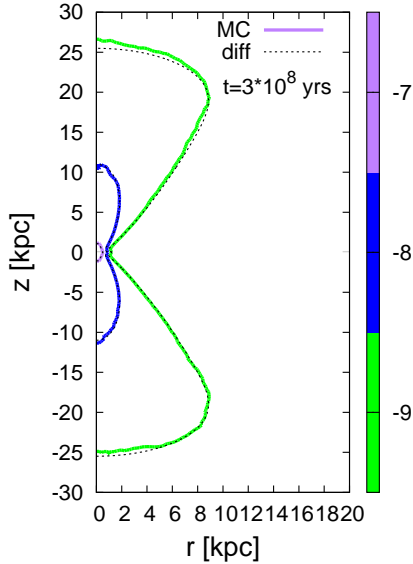}
\caption{A comparison plot showing the $log_{10}$ cosmic ray density contours (cm$^{-3}$) obtained
using both the Monte Carlo and differential equation solver methods. \lq\lq MC\rq\rq\/ 
refers to the Monte Carlo result and \lq\lq diff\rq\rq\/ to that from the differential equation 
solver. The different line colours indicate the corresponding contour value obtained from the \lq\lq MC\rq\rq\/ method, whose values are provided in the colour bar in the side-panel. The dashed lines are the results from the differential equation solution.}
\label{outflow_test}
\end{figure}

As a cross-check on the solutions to Equation~(\ref{diff_adv}) obtained,
a comparison of the results obtained from both the Monte Carlo and
differential equation solver methods are shown in Fig.~\ref{outflow_test}.
As is evident from this plot, very good agreement is found between
the two methods. Note the boundary and system setup conditions for this 
result were the same as that for the main paper text. Namely, a continuous 
source term, $Q_{CR}$  was located in the central region, a constant diffusion 
coefficient ($\mathcal{D}_{10\,{\rm GV}}$) was assumed, and an advective outflow 
described by Eq.~(\ref{EqnWind}) was adopted.

\section{Simulations for outflows at larger Galactocentric radii}

We show here some of the code verifications relating to the calculations 
presented in Section~\ref{local_outflow}. 

\begin{figure*}[t]
\includegraphics[width=0.49\textwidth]{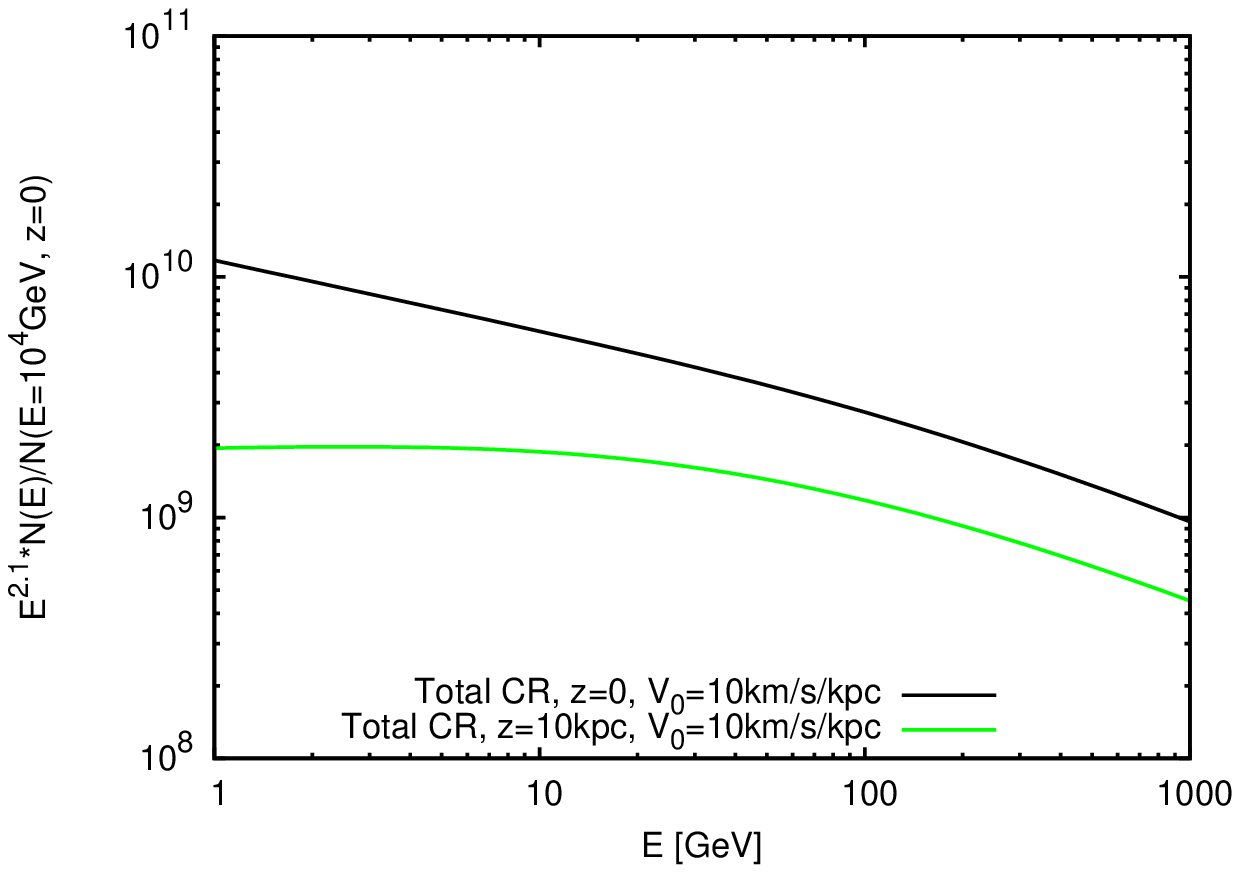}
\includegraphics[width=0.49\textwidth]{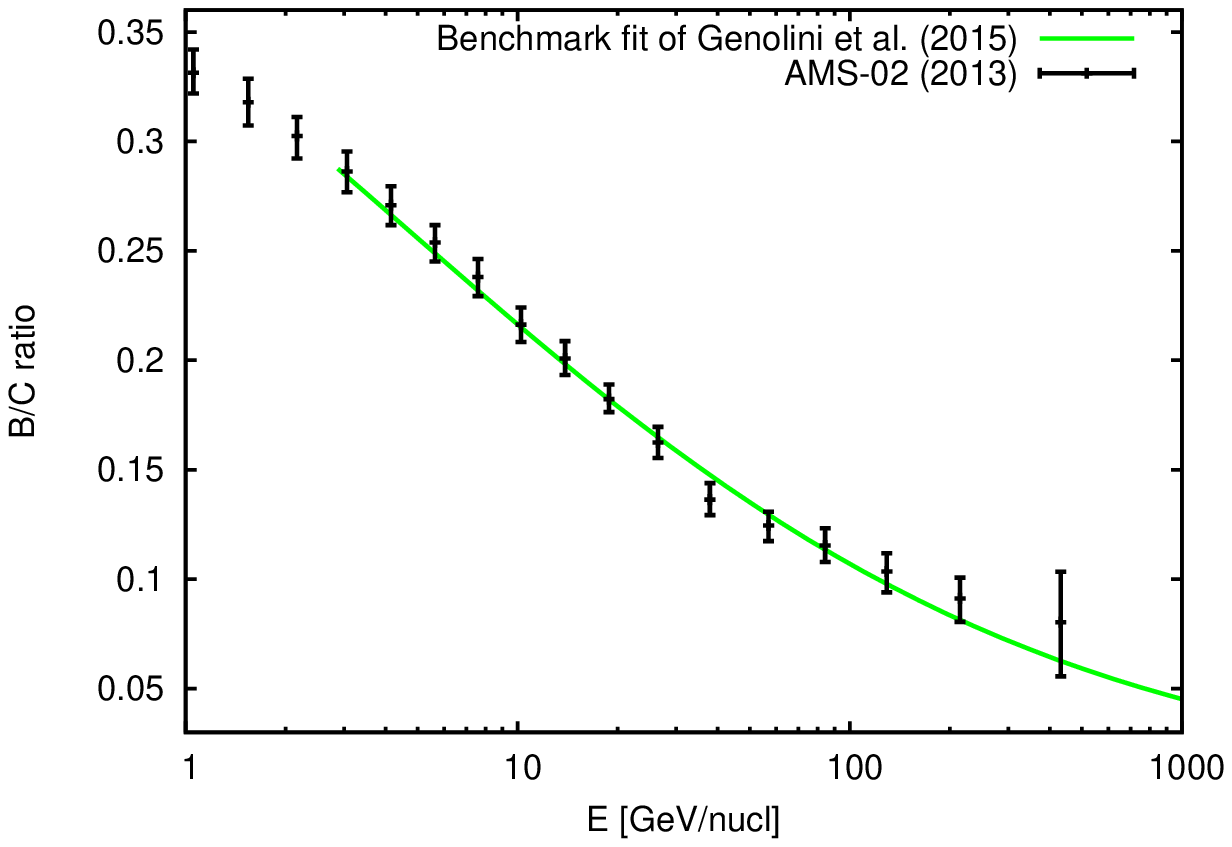}
\caption{{\it Left panel:} Calculation of the CR spectrum $E^{2.1}{\rm N}(E)/{\rm N}(E=10^4\,{\rm GeV},\,z=0)$ at $z=0$ (black solid line) and $z=10$\,kpc (green solid line) with our code, using the parameters of Figure~1 of Ref.~\cite{Bloemen1993} for $V_{0}=10$\,km\,s$^{-1}$\,kpc$^{-1}$, where $V(z)=3V_{0}z$. $H=20$\,kpc, $\mathcal{D}_{10\,{\rm GV}} \simeq 4.0 \times 10^{29}\,{\rm cm}^2\,{\rm s}^{-1}$, $\delta = 0.6$, and $\alpha = 2.1$. {\it Right panel:} Calculation of the boron-to-carbon ratio with our code, using the parameters of the \lq\lq benchmark fit\rq\rq\/ of Ref.~\cite{Genolini:2015cta} (green solid line) and compared with AMS-02 data~\cite{AMS2013} (black errorbars). The parameters of the \lq\lq benchmark fit\rq\rq\/ of~\cite{Genolini:2015cta} are: $V(z)=0$, $H=4$\,kpc, $\mathcal{D}_{10\,{\rm GV}} \simeq 4.8 \times 10^{28}\,{\rm cm}^2\,{\rm s}^{-1}$, and $\delta = 0.44$.}
\label{Appendix_B}
\end{figure*}

We verified that our code can reproduce the expected CR fluxes and spectra as 
functions of $z$ for the $V(z)=0$, $V(z)={\rm cst}$ and $V(z)\propto z$ 
wind profiles. As an example, we show in Fig.~\ref{Appendix_B} (left panel) our 
calculations of the normalized CR spectra multiplied by 
$E^{2.1}$, $E^{2.1}{\rm N}(E)/{\rm N}(E=10^4\,{\rm GeV},\,z=0)$, 
at $z=0$ (black solid line) and $z=10$\,kpc (green solid line), using the 
parameters of Figure~1 of Ref.~\cite{Bloemen1993} for $V_{0}=10$\,km\,s$^{-1}$\,kpc$^{-1}$, 
where
\begin{eqnarray}
  V(z) = 3V_{0}z\;.
\label{Eqn_App_B}
\end{eqnarray}
The parameters are: $H=20$\,kpc, 
$\mathcal{D}_{10\,{\rm GV}} \simeq 4.0 \times 10^{29}\,{\rm cm}^2\,{\rm s}^{-1}$, 
$\delta = 0.6$, and the spectral index at the sources is $\alpha = 2.1$. 
Good agreement is found between this result and its equivalent in Fig.~1 of 
Ref.~\cite{Bloemen1993}.

Concerning the calculation of the boron-to-carbon ratio, we verified, amongst other tests, 
that our code can reproduce the results for the \lq\lq benchmark fit\rq\rq\/ presented in Figure~3 of 
Ref.~\cite{Genolini:2015cta}. 
The parameters of this \lq\lq benchmark fit\rq\rq\/ are: $V(z)=0$, $H=4$\,kpc, 
$\mathcal{D}_{10\,{\rm GV}} \simeq 4.8 \times 10^{28}\,{\rm cm}^2\,{\rm s}^{-1}$, 
and $\delta = 0.44$. The results from our code are plotted in the right panel of 
Fig.~\ref{Appendix_B} (green solid line). The agreement with Fig.~3 of 
Ref.~\cite{Genolini:2015cta} is good.

\end{appendix}


\end{document}